# Towards mechanical characterization of soft digital materials for multimaterial 3D-printing

Viacheslav Slesarenko[1], Stephan Rudykh[2]

Department of Aerospace Engineering, Technion – Israel Institute of Technology

**Abstract**

We study mechanical behavior of soft rubber-like digital materials used in Polyjet multi-material 3D-printing to create deformable composite materials and flexible structures. These soft digital materials are frequently treated as linear elastic materials in the literature. However, our experiments clearly show that these materials exhibit significant non-linearities under large strain regime. Moreover, the materials demonstrate pronounced rate-dependent behavior. In particular, their instantaneous moduli as well as ultimate strain and stress significantly depend on the strain rate. To take into account both hyper- and viscoelasticity phenomena, we employ the Quasi-Linear Viscoelastic (QLV) model with instantaneous Yeoh strain-energy density function. We show that the QLV-Yeoh model accurately describes the mechanical behavior of the majority of the soft digital materials under uniaxial tension.

**Introduction**

Recently, additive manufacturing or 3D printing of architected materials has attracted significant attention [1–4]. Additive manufacturing, combined with rational design, allows one to create ultrastiff and ultralight materials [3], or materials, consisting of stiff components, but capable of recovering after large compressive deformation up to 50% [5]. Motivated by design and fabrication of reconfigurable and tunable materials and structures, soft microstructured materials capable of large deformations have become an active field of research. Architected materials with soft constituents demonstrate unique mechanical properties, such as negative Poisson's ratio [6], or act as machine-augmented composites, converting the applied force direction or absorbing elastic energy [7,8]. By varying the amount of soft constituents, together with their geometrical arrangements, it is possible to tailor the mechanical performance of the composites [9,10]. Moreover, such composites may be used to actively and reversibly control wave propagation via deformations and elastic instabilities [11,12]. Though an addition of soft components in architected materials may be beneficial for achieving new functionalities, it requires more complicated material fabrication techniques that allow producing soft-stiff deformable composites. Recently, new UV-curing materials have been introduced into multimaterial 3D printing; the developed by Stratasys rubber-like materials show the ability of ustaining extremely large elastic deformations before failure at around 200-300 % strain level. These soft UV curing resins are used in the Polyjet printing process. During Polyjet 3D-printing, small droplets of photopolymers are deposited onto the horizontal platform through the series of inkjet printing heads. After deposition of each layer, the ultraviolet lamps cure the photopolymer. This Polyjet 3D-printing technique allows fabrication of multimaterial structures with the resolution as fine as 16 micron with up to three different materials in printed models. Furthermore, by using different nozzles for different materials, it is possible to locally mix several materials and, therefore, to obtain homogeneous soft resins with various mechanical properties. In particular, by mixing soft TangoPlus (TP) polymer with stiff VeroWhite (VW) polymer, it is possible to obtain the so-called digital materials (DMs) with different mechanical properties.

---

[1] E-mail: sl.slesarenko@gmail.com
[2] E-mail: rudykh@technion.ac.il

The DMs have been employed to develop the concept of flexible armor [13], or to mimic the natural structures of nacre [14], suture interfaces [15,16] and other biological materials [17]. DMs have also been used to prototype membranes [18], arteries [19] et al. Surprisingly, the reported mechanical properties of the materials under study vary significantly in the literature. Table 1 summarizes the mechanical properties of TangoPlus material as reported in different studies. Note the significant variations of the initial shear modulus of the digital material. Recent works have revealed that certain variation of the shear modulus can be caused by the printing process anisotropy [20,21]. However, the reported anisotropy of the mechanical properties does not exceed 20% [9,22]. Additional factors that may affect mechanical performance include temperature, humidity or illumination during printing as well as during testing [20]. The DMs under study have glass transition temperatures in the -10 to +58 °C range [23], therefore, small temperature deviations during UV-curing or testing may affect measured mechanical properties. The properties of DMs can also change with time, due to aging occurring in photopolymers. Therefore, time interval from manufacturing to testing can cause some alteration in the properties of the materials [20]. Summarizing, we would like to emphasize that full characterization of the DMs is a challenging task, and most studies of composite materials require preliminary testing of the constituents under specific conditions of 3D printing and testing. Thus, UV-curing soft resins and DMs require specific attention and careful consideration of their complicated mechanical behavior.

We would like to note that soft DMs exhibit significant nonlinear mechanical behavior, which is frequently neglected in the available literature – we summarize reported mechanical models in Table 1. For example, these DMs are frequently considered as linear elastic materials. Linear elastic model can provide adequate approximations for material behavior for the small strains, typically not exceeding 5%; application of linear elastic model under the large strain regime may, however, lead to inaccurate results and conclusions. Another simplification of the mechanical behavior occurs when their mechanical response is considered to be independent of the strain rate. The effects of viscoelasticity on composite behavior have been analyzed in [21,24,25], showing significant influence of rate-dependent behavior on the overall response of the composites. Rate-dependent behavior of the DMs drastically increases the complexity of the constitutive models, but however, it is of importance for accurate modeling of the materials and predictions of their behavior. For example, the studies by [24,26] showed that viscoelasticity leads to an increased tunability of wrinkling patterns, occurring in the soft laminates.

**Table 1**. Elastic modulus of TangoPlus/TangoBlackPlus materials, reported in different studies. LE – linear elastic model, NH – neo-Hookean model, AB – Arruda-Boyce model.

| Reference | Constitutive Model | Reported modulus (Y – Young's S – shear) , [MPa] | Poisson's Ratio | Shear modulus[3] [MPa] |
|---|---|---|---|---|
| [9] | LE | 0.47 (Y) | - | 0.158 |
| [10] | NH | 0.56 (Y) | 0.49 | 0.188 |
| [24] | NH | 0.2 (S) | - | 0.200 |
| [14] | NH | 0.21 (S) | - | 0.210 |
| [16] | LE | 0.63 (Y) | - | 0.211 |
| [27] | AB | 0.213 (S) | - | 0.213 |
| [13] | - | 0.78 (Y) | - | 0.261 |
| [28] | LE | 0.9 (Y) | - | 0.302 |
| [22] | LE | 0.99 (Y) | 0.48 | 0.330 |

Rate dependency and non-linearities due to large deformations are the essential aspects of rich mechanical behavior of the soft DMs used in multimaterial 3D-printing. Here we address these important factors in our attempt to describe mechanical behavior of the DMs using the Quasi-Linear Viscoelastic (QLV) model, which combines hyper- and viscoelasticity phenomena, and to discuss its features and limitations. We note that the paper objective is not the attempt to provide full characterization of the DMs in the complexity of their behavior but exploration of possible ways of accounting for some significant aspects of their mechanical behavior, such as material non-linearity and rate sensitivity.

**Theoretical background**

Modeling of rate-dependent materials undergoing finite deformations is a well-known problem that attracted significant attention (see for example the review by Wineman [29] and references therein). The stress response for a viscoelastic non-linear material can be expressed as

$$\boldsymbol{\sigma}(t) = \boldsymbol{F}(t)\mathcal{G}[\boldsymbol{C}(t-s)|_{s=0}^{\infty}]\boldsymbol{F}(t)^T, \qquad (1)$$

---

[3] For comparison reasons, the initial shear modulus is calculated for LE model as $\mu = 0.5\, E/(1+\nu)$. If Poisson's ratio is not reported, we consider the material to be nearly incompressible with $\nu = 0.49$ in the calculation of the initial shear modulus. If the anisotropy of the elastic modulus is reported, we calculate the average elastic modulus.

where $\boldsymbol{\sigma}(t)$ is the Cauchy stress tensor, $\boldsymbol{F}$ is the deformation gradient, $\boldsymbol{C} = \boldsymbol{F}^T\boldsymbol{F}$ is the right Cauchy-Green tensor, and $\mathcal{G}$ is the response potential capturing the loading history. Adopting the Pipkin-Rogers constitutive model [29,30] of the response potential and assuming that deformation starts at $t = 0$, we write the expression for Cauchy stress tensor as

$$\boldsymbol{\sigma}(t) = \boldsymbol{F}(t)\left\{\boldsymbol{K}[\boldsymbol{C}(t),0] + \int_0^t \frac{\partial}{\partial(t-s)}\boldsymbol{K}[\boldsymbol{C}(s),t-s]ds\right\}\boldsymbol{F}(t)^T, \qquad (2)$$

where $\boldsymbol{K}$ is the kernel, responsible for the description of the fading memory. Under material incompressibility assumption, the Jacobian $J = \det \boldsymbol{F}(t) = \det \boldsymbol{C}(s) = 1$, and Eq.(2) modifies as

$$\boldsymbol{\sigma}(t) = -p\boldsymbol{I} + \boldsymbol{F}(t)\left\{\boldsymbol{K}[\boldsymbol{C}(t),0] + \int_0^t \frac{\partial}{\partial(t-s)}\boldsymbol{K}[\boldsymbol{C}(s),t-s]ds\right\}\boldsymbol{F}(t)^T, \qquad (3)$$

where $p$ is an unknown Lagrange multiplayer, and $\boldsymbol{I}$ is the identity tensor. As one can see from Eq. (3), the kernel $\boldsymbol{K}$ depends on time and right Cauchy-Green tensor. A special case of this model, when $\boldsymbol{K}(\boldsymbol{C},s)$ can be decomposed as $\boldsymbol{K}(\boldsymbol{C},s) = \boldsymbol{S}^e[\boldsymbol{C}]G(s)$, corresponds to the well-known Quasi-Linear Viscoelastic (QLV) model, which is widely used, for instance, in modelling of soft tissues. Thus, Cauchy stress tensor takes the form

$$\boldsymbol{\sigma}(t) = -p\boldsymbol{I} + \boldsymbol{F}(t)\left\{\boldsymbol{S}^e[\boldsymbol{C}(t)] + \int_0^t \boldsymbol{S}^e[\boldsymbol{C}(s)]\frac{\partial G(t-s)}{\partial(t-s)}ds\right\}\boldsymbol{F}(t)^T. \qquad (4)$$

The term $\boldsymbol{S}^e[\boldsymbol{C}(t)]$ can be considered as the effective (instantaneous) second Piola-Kirchhoff elastic stress tensor [31,32]. After decomposition of $\boldsymbol{S}^e[\boldsymbol{C}(t)]$ into two terms, corresponding to the second Piola-Kirchhoff stress of the deviatoric and hydrostatic Cauchy stress components, Cauchy stress tensor is expressed as [31]

$$\boldsymbol{\sigma}(t) = J^{-1}\boldsymbol{F}(t)\left[\boldsymbol{S}_D^e(t) + \int_0^t D'(t-s)\boldsymbol{S}_D^e(s)ds\right]\boldsymbol{F}^T(t)$$
$$+ J^{-1}\boldsymbol{F}(t)\left[\boldsymbol{S}_H^e(t) + \int_0^t H'(t-s)\boldsymbol{S}_H^e(s)ds\right]\boldsymbol{F}^T(t), \qquad (5)$$

where $D(t)$ and $H(t)$ are the relaxation functions, and the effective elastic second Piola-Kirchhoff stress tensors $\boldsymbol{S}_D^e$ and $\boldsymbol{S}_H^e$ are given by

$$\boldsymbol{S}_D^e = 2\left[\frac{1}{3}(I_2 W_2 - I_1 W_1)\boldsymbol{C}^{-1} + W_1\boldsymbol{I} - I_3 W_2 \boldsymbol{C}^{-2}\right] \qquad (6)$$

and

$$\boldsymbol{S}_H^e = 2\left[\frac{2}{3}I_2 W_2 + \frac{1}{3}I_1 W_1 + I_3 W_3\right]\boldsymbol{C}^{-1}. \qquad (7)$$

In Eqs. (6) and (7), $I_1, I_2, I_3$ denote the corresponding invariants of the deformation gradient, and

$$W_k = \frac{\partial W}{\partial I_k}, \quad k = 1,2,3. \tag{8}$$

If the material is assumed to be incompressible, the second term in (5) is substituted by $-p(t)\boldsymbol{I}$, where $p(t)$ is Lagrange multiplier, and, hence

$$\boldsymbol{\sigma}(t) = \boldsymbol{F}(t)\left[\boldsymbol{S}_D^e(t) + \int_0^t D'(t-s)\boldsymbol{S}_D^e(s)ds\right]\boldsymbol{F}^T(t) - p\boldsymbol{I} \tag{9}$$

In order to employ the QLV model, the rate-independent strain-energy density function needs to be specified. This rate-independent strain-energy density function reflects the instantaneous response of the material. Here, we consider the classical *incompressible* two-term Yeoh model [33], for which the strain-energy density function is defined as

$$W = \frac{\mu}{2}\left[(I_1 - 3) + \frac{\alpha}{2}(I_1 - 3)^2\right]. \tag{10}$$

Note that for $\alpha = 0$ this model reduces to neo-Hookean model.

For Yeoh model, we obtain the following derivatives of the energy density function

$$W_1 = \frac{\mu}{2}(1 - 3\alpha + \alpha I_1), \quad W_2 = 0, \quad W_3 = 0 \tag{11}$$

Therefore, from (6)

$$\boldsymbol{S}_D^e = -\frac{\mu}{3}[\alpha I_1^2 + (1-3\alpha)I_1]\boldsymbol{C}^{-1} + \mu(1 - 3\alpha I_1)\boldsymbol{I}. \tag{12}$$

Let us consider the mechanical response of the QLV material, subjected to uniaxial tension. Assuming that tension is applied in the direction along the basis vector $\boldsymbol{e}_1$, the corresponding deformation gradient $\boldsymbol{F}(t)$ takes form

$$\boldsymbol{F}(t) = \lambda(t)\boldsymbol{e}_1\otimes\boldsymbol{e}_1 + \lambda_2(t)\boldsymbol{e}_2\otimes\boldsymbol{e}_2 + \lambda_2(t)\boldsymbol{e}_3\otimes\boldsymbol{e}_3, \tag{13}$$

where $\lambda_2 = \lambda^{-1/2}$ due to the incompressibility assumption.

From (12) together with (13) we obtain the following expressions for the components of the instantaneous second Piola-Kirchhoff stress tensor

$$S_{D11}^e = \frac{2}{3}\mu[\alpha\lambda^2 + (1 - 3\alpha) + \alpha\lambda^{-1} - (1-3\alpha)\lambda^{-3} - 2\alpha\lambda^{-4}] \tag{14}$$

and

$$S_{D22}^e = -\frac{\mu}{3}[\alpha\lambda^5 + (1-3\alpha)\lambda^3 + \alpha\lambda^2 - (1-3\alpha) - 2\alpha\lambda^{-1}]. \tag{15}$$

Therefore the expressions for the components of Cauchy stresses (9) take form

$$\sigma_{11}(t) = \lambda^2(t)[S_{D11}^e(t) + \int_0^t D'(t-s)S_{D11}^e(s)ds] - p \tag{16}$$

and

$$\sigma_{22}(t) = \lambda^{-1}(t)\left[S_{D22}^e(t) + \int_0^t D'(t-s)S_{D22}^e(s)ds\right] - p. \tag{17}$$

The stress free boundary condition implies that $T_{22}(t) = 0$, from which the unknown $p$ is found, thus, leading to the following expression for the Cauchy stress component in the tension direction

$$\begin{aligned}\sigma_{11}(t) = {} & \mu[\alpha\lambda^3(t) + (1-3\alpha)\lambda(t) + 2\alpha][\lambda(t) - \lambda^{-2}(t)] \\ & + \mu \int_0^t D'(t-s)[\alpha\lambda^3(s) + (1-3\alpha)\lambda(s) + 2\alpha][\frac{2}{3}\lambda^2(t)\{\lambda^{-1}(s) \\ & - \lambda^{-4}(s)\} + \frac{1}{3}\lambda^{-1}(t)\{\lambda^2(s) - \lambda^{-1}(s)\}]ds\end{aligned} \tag{18}$$

For neo-Hookean QLV material Eq. (18) simplifies, and it takes the form

$$\begin{aligned}\sigma_{11}(t) = {} & \mu[\lambda^2(t) - \lambda^{-1}(t)] \\ & + \mu \int_0^t D'(t-s)\left[\frac{2}{3}\lambda^2(t)\{\lambda(s) - \lambda^{-3}(s)\} \right. \\ & \left. + \frac{1}{3}\lambda^{-1}(t)\{\lambda^3(s) - 1\}\right]ds.\end{aligned} \tag{19}$$

**Experimental study**

Material parameters of rate-dependent materials can be derived from different sets of experiments, including stress relaxation, creep, uniaxial tension, etc. However, uniaxial tension is one of the most common modes of deformation, which motivates us to use this mode to analyze the mechanical behavior of the DMs used in three-dimensional printing. The studied DMs are produced by local mixing of the two polymers, namely stiff VeroWhite (VW) and soft TangoPlus (TP). While the manufacturer does not provide the mixing ratios of these two polymers for producing DMs with different property index, these ratios are estimated by the material consumption during 3D-printing process. Following the notation of Stratasys, we use Shore A index to name the corresponding materials; for instance, DM with Shore index 40 is referred as DM40. Figure 1 summarizes the approximate composition of DMs with different Shore index.

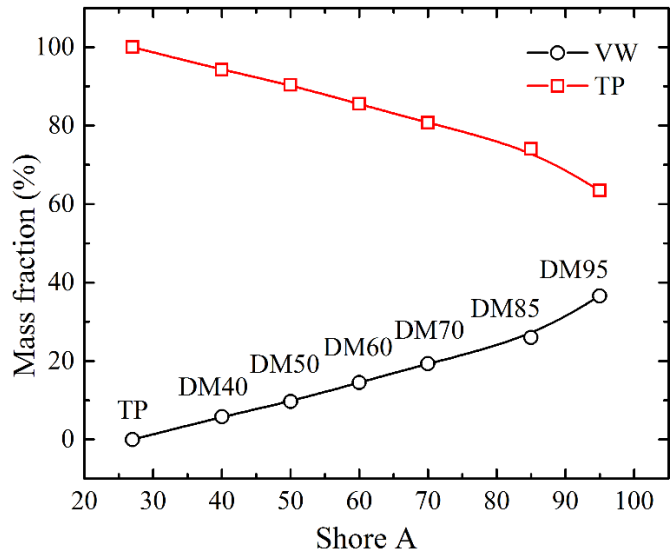

**Figure 1**. The approximate composition of DMs with different Shore A index

The "dog-bone" samples of different DMs with geometry, corresponding to Type IV ASTM D638 standard, have been fabricated by the use of Objet Connex 260 3D-printer. All specimens have been printed with the same orientation in order to avoid the anisotropy effects that may be induced during 3D-printing. For each digital material at least three samples have been printed. Then the samples have been subjected to uniaxial tension at different strain rates. In order to avoid the variation of testing conditions and possible aging process, fabrication and testing of all specimens have been performed within two weeks period, and the time interval between fabrication of the sample and its testing were not less than 24 hours and did not exceed 48 hours in order to completely finish the curing process and to exclude subsequent aging effects. The tensile strain and strain rate in gauge part of the specimens have been determined with the help of a video-extensometer. The applied strain rates were equal to $1.2 \cdot 10^{-1}$, $1.2 \cdot 10^{-2}$, $1.2 \cdot 10^{-3}$ s$^{-1}$, corresponding to quasi-static loadings with intermediate strain rates, typically used in the experiments reported in the literature. In order to estimate the possible experimental error, additional four samples of DM50 and DM95 have been printed and tested. The obtained stress-strain curves illustrate a good repeatability of the experiments (see Figure 2d).

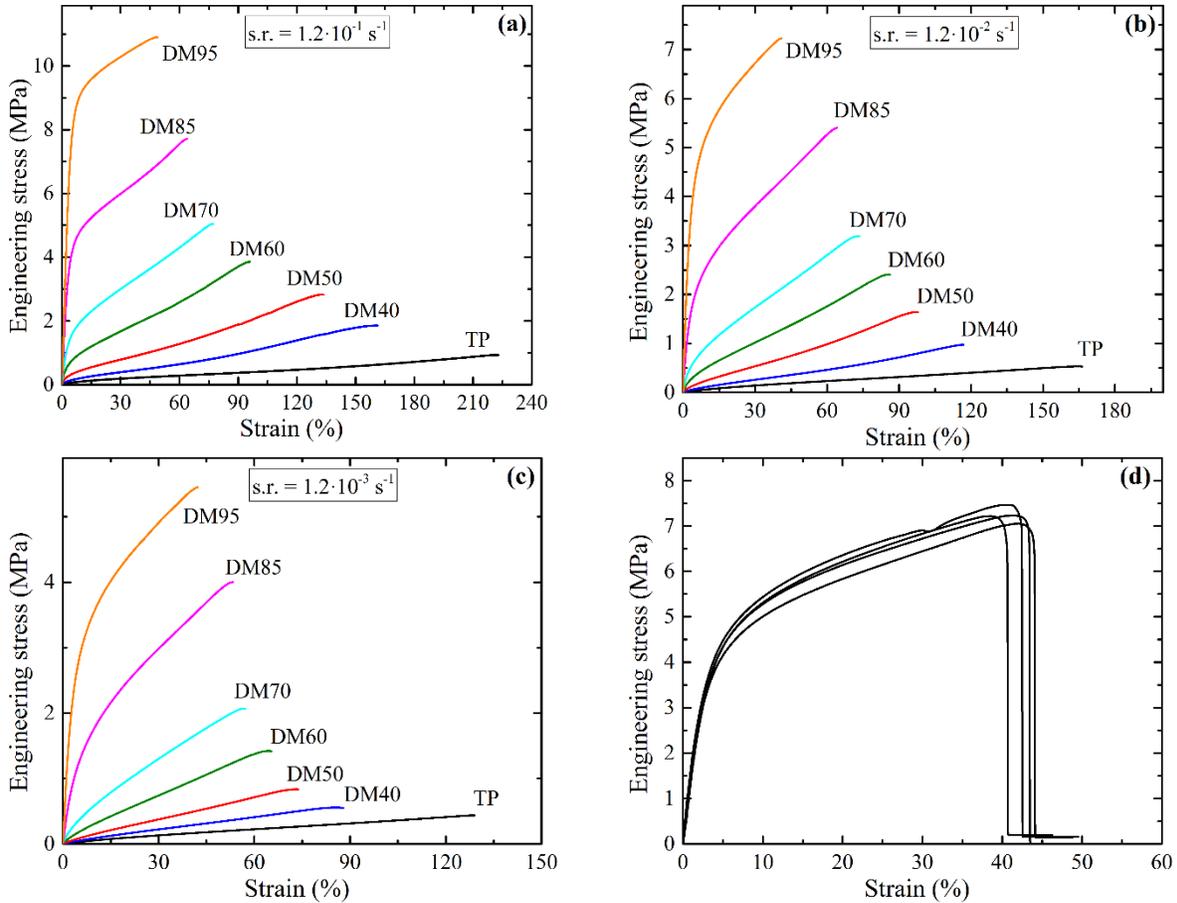

**Figure 2.** The stress-strain curves obtained for uniaxial tension of DMs with strain rates $1.2 \cdot 10^{-1}$ (a), $1.2 \cdot 10^{-2}$ (b) and $1.2 \cdot 10^{-3}$ s$^{-1}$ (c). Comparison of the four tests, carried out on the DM95 material with strain rate of $1.2 \cdot 10^{-2}$ s$^{-1}$ (d)

Figure 2 shows the stress-strain curves, obtained during uniaxial tension of DMs with strain rates of $1.2 \cdot 10^{-1}$ (a), $1.2 \cdot 10^{-2}$ (b) and $1.2 \cdot 10^{-3}$ s$^{-1}$ (c). While the stress-strain curves, observed for the DMs with low mass fraction of VW, such as TP, DM40, DM50 are close to linear, especially under slow loadings, the mechanical responses of the DMs with larger mass fraction of the stiff VW polymer are highly nonlinear. One may see that the materials with lower mass fraction of VW demonstrate more compliant behavior and higher elongation till failure. Figure 3a shows the dependency of the ultimate strain on mass fraction of VW for different strain rates. One may see that the ultimate strain decreases with an increase in the mass fraction of VW. The opposite trend is observed for the ultimate stress. An increase in the mass fraction of VW leads to an increase in the ultimate stress (Figure 3b). For instance, while the pure TP can be elongated up to 265 % without failure and can withhold the stress as low as 0.5 MPa, DM95 demonstrates much higher ultimate stress of 7.2 MPa, but fails after elongation of only 140 %. Moreover, for all considered DMs ultimate strain increases with an increase in the strain rate. For example, the ultimate strain, achievable by TP under fast loading with the strain rate of $1.2 \cdot 10^{-1}$ s$^{-1}$ is 1.5 times larger than for the hundred times slower loading. From Figure 3a and b one can conclude that for faster loading the specimen of DMs can reach higher strains and stresses without failure.

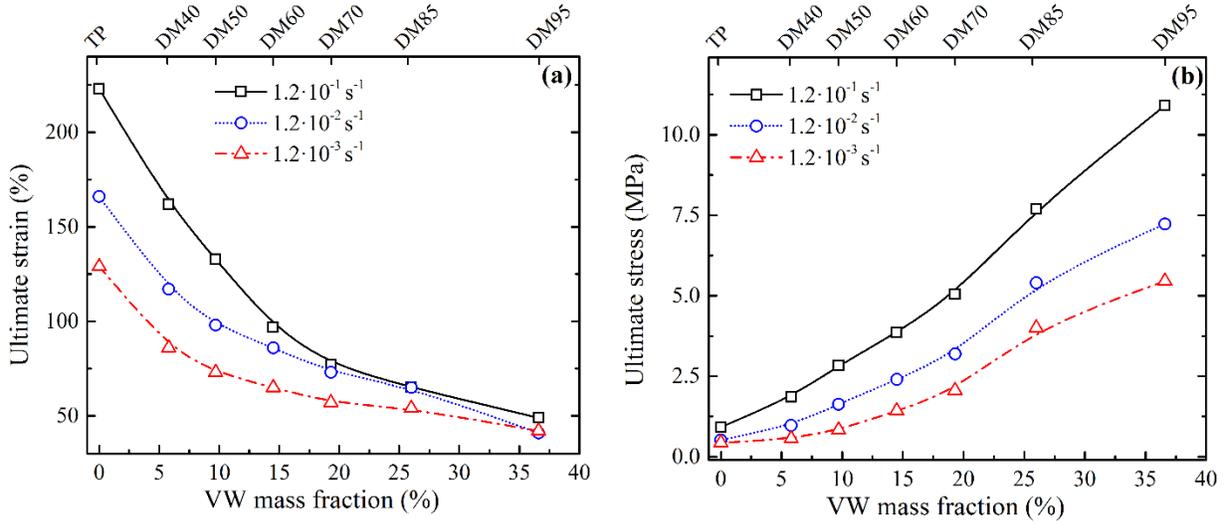

**Figure 3.** The dependencies of ultimate strain (a) and ultimate stress (b) on VW mass fraction.

**Application of QLV model to the experimental results**

The comparison of the stress-strain curves, obtained for the same material, but for the tension with different strain rates, allows us to more clearly demonstrate the viscoelastic behavior of the soft materials. For instance, Figure 4 shows the corresponding stress-strain curves for the DM40. Typically for viscoelastic materials, the faster loading leads to a stiffer response of the material in comparison with slower loading. The relation between stress and strain in viscoelastic material cannot be represented by only two elastic constants even for infinitesimal strain theory. In QLV model, the influence of the loading history on the current stress state is defined through the stress relaxation function $D(t)$ (see Eq. (9)). Here we use the classical Prony series, namely

$$D(t) = 1 - \sum_{i=1}^{n} \gamma_i (1 - e^{-\frac{t}{\tau_i}}) \qquad (20)$$

where $\tau_i$ and $\gamma_i$ are the corresponding relaxation times and relaxation coefficients, respectively.

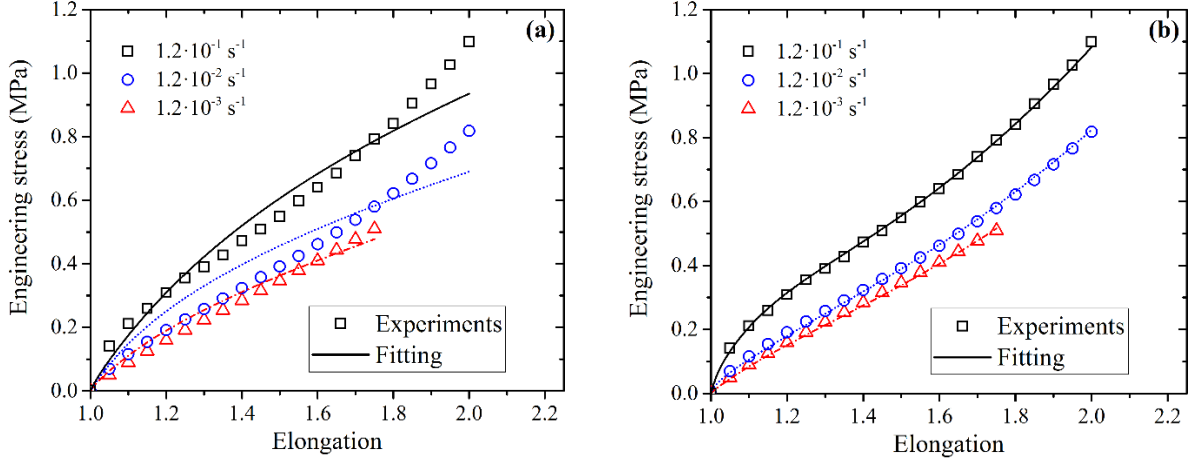

**Figure 4.** Fitting of the experimental data obtained for DM40 by using QLV models with neo-Hookean (a) and Yeoh (b) instantaneous strain-energy density functions.

Since relaxation processes occur at different timescales, one Prony series is usually insufficient to accurately capture the material stress response. Therefore, here we use three Prony series with relaxation times $\tau = 0.1,\ 1$ and $100\ s$. The observed stress-strain curves, obtained for different strain rates, have been fitted by using Levenberg-Marquardt method with the least squares criterion. Figure 4a and b show the fitting curves, obtained under the assumption of QLV model with neo-Hookean (19) and Yeoh (18) instantaneous strain-energy density functions, respectively. One may see, that the model with neo-Hookean instantaneous strain-energy density function inadequately describes the behavior of the DM for deformation larger than about 10%, while the QLV model with Yeoh instantaneous response provides quite remarkable agreement with the experimental data. The limitations of neo-Hookean model to capture the mechanical response of some nonlinear materials is well known; the addition of only one term in the strain-energy density function, drastically improves the accuracy of the QLV model for the DMs.

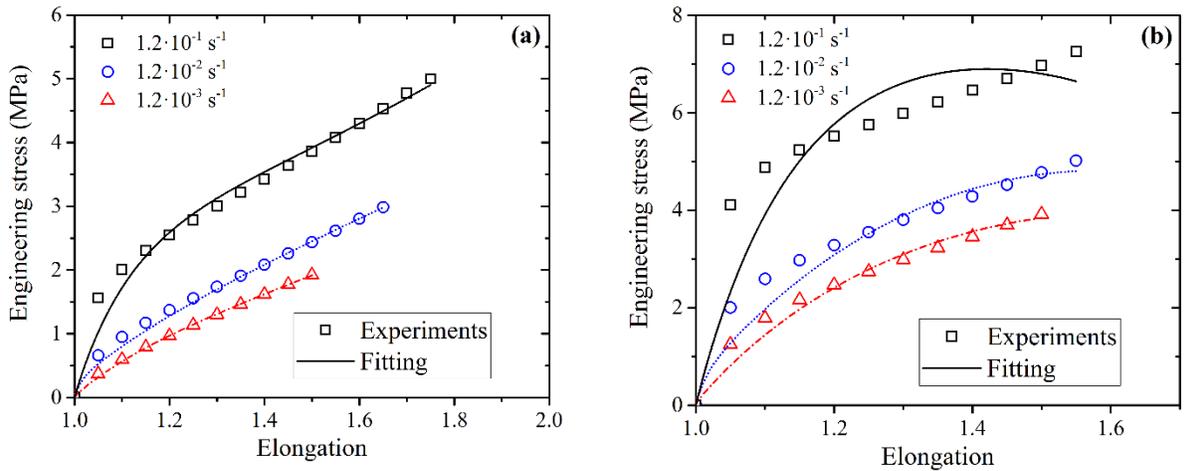

**Figure 5.** Fitting of the experimental data obtained for DM70 (a) and DM85 (b) by using the QLV model with Yeoh instantaneous strain-energy density function.

Figure 5 shows the fitting results for two different DMs. The QLV model with instantaneous response defined by Yeoh strain-energy density function, capture very well the behavior of soft DMs with Shore index not exceeding 70 (Figure 5a). The corresponding material constants are summarized in Table 2.

**Table 2.** Material constants of DMs for the QLV model with instantaneous Yeoh strain-energy density function. $\mu_l$ represents the value of the long-term shear modulus, characterizing the material response under the infinitely slow loading. The relaxation times are $\tau_1 = 0.1\ s, \tau_2 = 1\ s,\ \tau_3 = 100\ s$.

| Material | $\mu$ (MPa) | $\alpha$ | $\gamma_1$ | $\gamma_2$ | $\gamma_3$ | $\mu_l$ (MPa) |
|---|---|---|---|---|---|---|
| TP | 0.688 | 0.060 | 0.393 | 0.348 | 0.009 | 0.17 |
| DM40 | 1.626 | 0.265 | 0.499 | 0.297 | 0.04 | 0.27 |
| DM50 | 2.103 | 0.218 | 0.258 | 0.376 | 0.105 | 0.55 |
| DM60 | 4.77 | 0.188 | 0.173 | 0.523 | 0.112 | 0.91 |
| DM70 | 8.53 | 0.088 | 0.079 | 0.635 | 0.094 | 1.64 |

The obtained material parameters (summarized in Table 2) provide a plausible explanation for the inconsistency in the elastic modulus of TP reported in the literature (as summarized in Table 1). Depending on the strain rate, the shear modulus, estimated from the uniaxial tension/compression experiments under the assumption of rate-independent behavior, can take values in the range from 0.17 to 0.69 MPa. As expected, the instantaneous shear modulus, $\mu$ as well as the long-term shear modulus, $\mu_l$ increase with an increase in the mass fraction of VW in the DMs. At the same time, by comparing the relaxation coefficients $\gamma$ for the relaxation times of 0.1 and 1 s, one may notice a more pronounced contribution of the slower relaxation processes in the stiffer DMs with higher mass fraction of VW. For instance, while for DM40 the Prony series with relaxation time 0.1 s provide a "decrease" in the shear modulus almost in two times, the processes, acting at the same time scale in the DM70, result in only 7.9% drop in the value of the shear modulus. Meanwhile, the contribution of the slower relaxation processes increases from 0.297 to 0.635. Remarkably, the rate-dependent behavior of the pure TP does not follow this observation. Presumably, the relaxation mechanisms, occurring in the DMs, produced as the mix of TP and VW, slightly differ from the deformation mechanisms in the pure TP. Summarizing the results, presented in Table 2, we can make the observation that, from a practical point of view, the rate-dependent behavior of the softer DMs may be neglected for the faster loading in comparison with DMs with higher shear moduli.

We note that in our work DMs have been assumed to be incompressible (or nearly incompressible), which is in a good agreement with the behavior observed in experiments and reported in the literature. This incompressibility assumption significantly simplifies calculations, leading to vanishing of the second term in Eq. (5), capturing the relaxation of volumetric part. Unfortunately, even incompressible QLV model with instantaneous Yeoh strain-energy density function does not provide an adequate description of the behavior of the two stiffest DMs, namely DM85 and DM95, as shown in Figure 5b. The main inconsistency occurs at the very beginning of the loading and the model is not able to capture the difference in the response at the different strain rates even for the strain level not exceeding 7%. We note that the inconsistency with experimental data has not been resolved even when compressibility has been taken onto account. In Appendix, we discuss the influence of the compressibility on the stress-strain response, linearizing the QLV model for infinitesimal strain theory. Therefore, we can assume that QLV model (Eq. (4)) is not able to accurately capture the rate-dependent behavior of DM85 and DM95, and more rigorous models (such as, for example, models without decomposition of relaxation kernel term) are required. At the same time, we have showed that the QLV model with incompressible instantaneous Yeoh strain-energy density function, accurately describes the complicated mechanical behavior of the DMs with Shore index not exceeding 70.

**Conclusions**

We have studied the mechanical behavior of soft rubber-like DMs used in multimaterial 3D-printing. Our experimental results show the nonlinear behavior and essential dependence of DMs mechanical properties on the applied strain rate. Moreover, we have found that the ultimate strain of the DMs is significantly influenced by the strain rate. Thus, for example, the ultimate strain of TP is more than 1.5 times higher for the loading with strain rate of $1.2 \cdot 10^{-1}$ $s^{-1}$ in comparison with a slower loading of $1.2 \cdot 10^{-3}$ $s^{-1}$. In order to make allowance for viscoelastic effects, the QLV model has been employed to fit the experimental data. We have observed that the QLV model with instantaneous neo-Hookean strain-energy density function does not describe the behavior of the DMs under large strain levels exceeding 10-15%. However, the QLV model with instantaneous Yeoh strain-energy density function (as compared to the QLV-neo-Hookean model) is capable of capturing the experimental data with good accuracy. However, even the use of the QLV-Yeoh model does not allow us to properly describe the behavior of the DMs with larger amounts of stiff VW polymer in the mixture (DM85 and DM95). The inconsistency is highlighted and discussed in the Appendix. We conclude by noting that modeling of the DMs requires appropriate constitutive models, which are capable of describing rich mechanical behavior of the soft DMs. Application of such models can help computationally predict and discover the new phenomena in complex functional systems realizable by multimaterial 3D-printing.

**Appendix**

Here we discuss the possible influence of compressibility on the stress-strain response of viscoelastic soft materials. Since the derivation of the stress-strain relation for compressible QLV material model is an involving task that requires numerical calculations, we consider the case of infinitesimal strain. The analogue of the Eq. (5) under the assumption that the deformation starts at $t = 0$ for the infinitesimal strain theory has the following form [31]

$$\boldsymbol{\sigma}(t) = 2\mu \left[\boldsymbol{\varepsilon}(t) - \frac{1}{3}(tr\boldsymbol{\varepsilon}(t))\right] \boldsymbol{I} + 2\mu \int_0^t D'(t-s) \left[\boldsymbol{\varepsilon}(s) - \frac{1}{3}(tr\boldsymbol{\varepsilon}(s))\boldsymbol{I}\right] ds \\ + \kappa\, tr\boldsymbol{\varepsilon}(t)\boldsymbol{I} + \kappa \int_0^t H'(t-s) tr\boldsymbol{\varepsilon}(s)\boldsymbol{I} ds, \quad (21)$$

where $\boldsymbol{\sigma}$ and $\boldsymbol{\varepsilon}$ are the second-order stress and strain tensors, respectively; $\mu$ and $\kappa$ are the instantaneous shear and bulk elastic moduli.

For uniaxial tension, we can obtain the following expressions for the $\sigma_{11}$ and $\sigma_{22}$ components of the stress tensor

$$\sigma_{11}(t) = \varepsilon_{11}(t)\left[\kappa + \frac{4\mu}{3}\right] \\ + \int_0^t \left[\kappa H'(t-s) + \frac{4\mu}{3} D'(t-s)\right] \varepsilon_{11}(s) ds + \varepsilon_{22}(t)\left[2\kappa - \frac{4\mu}{3}\right] \\ + \int_0^t \left[2\kappa H'(t-s) - \frac{4\mu}{3} D'(t-s)\right] \varepsilon_{22}(s) ds \quad (22)$$

and

$$\sigma_{22}(t) = \varepsilon_{11}(t)\left[\kappa - \frac{2\mu}{3}\right] \\ + \int_0^t \left[\kappa H'(t-s) - \frac{2\mu}{3} D'(t-s)\right] \varepsilon_{11}(s) ds + \varepsilon_{22}(t)\left[2\kappa + \frac{2\mu}{3}\right] \\ + \int_0^t \left[2\kappa H'(t-s) + \frac{2\mu}{3} D'(t-s)\right] \varepsilon_{22}(s) ds. \quad (23)$$

Assuming that loading occurs with the constant strain rate, and utilizing the boundary condition $\sigma_{22}(t) = 0$, for some cases one may express $\varepsilon_{11}(t)$ through $\varepsilon_{22}(t)$ from Eq. (23) and obtain an expression for $\sigma_{11}(t)$. Typically, these calculations are rather complex and are performed numerically. However, let us consider the case when the relaxation functions for volumetric and deviatoric parts of strain tensor coincide, in other words, $D(t) = H(t)$ for any $t$. Thus, one can show that the closed form expression for $\sigma_{11}(t)$ is

$$\sigma_{11}(t) = \mu \frac{9q}{3q+1}\left[\varepsilon_{11}(t) + \int_0^t D'(t-s)\varepsilon_{11}(s)ds\right] \quad (24)$$

where $q = \kappa/\mu$ is the contrast between bulk and shear moduli for instantaneous loading. One may see, that when $q \to \infty$ (the Poisson ratio $\nu \to 0.5$), the material becomes incompressible and therefore

$$\sigma_{11}(t) \to 3\mu\left[\varepsilon_{11}(t) + \int_0^t D'(t-s)\varepsilon_{11}(s)ds\right] \quad (25)$$

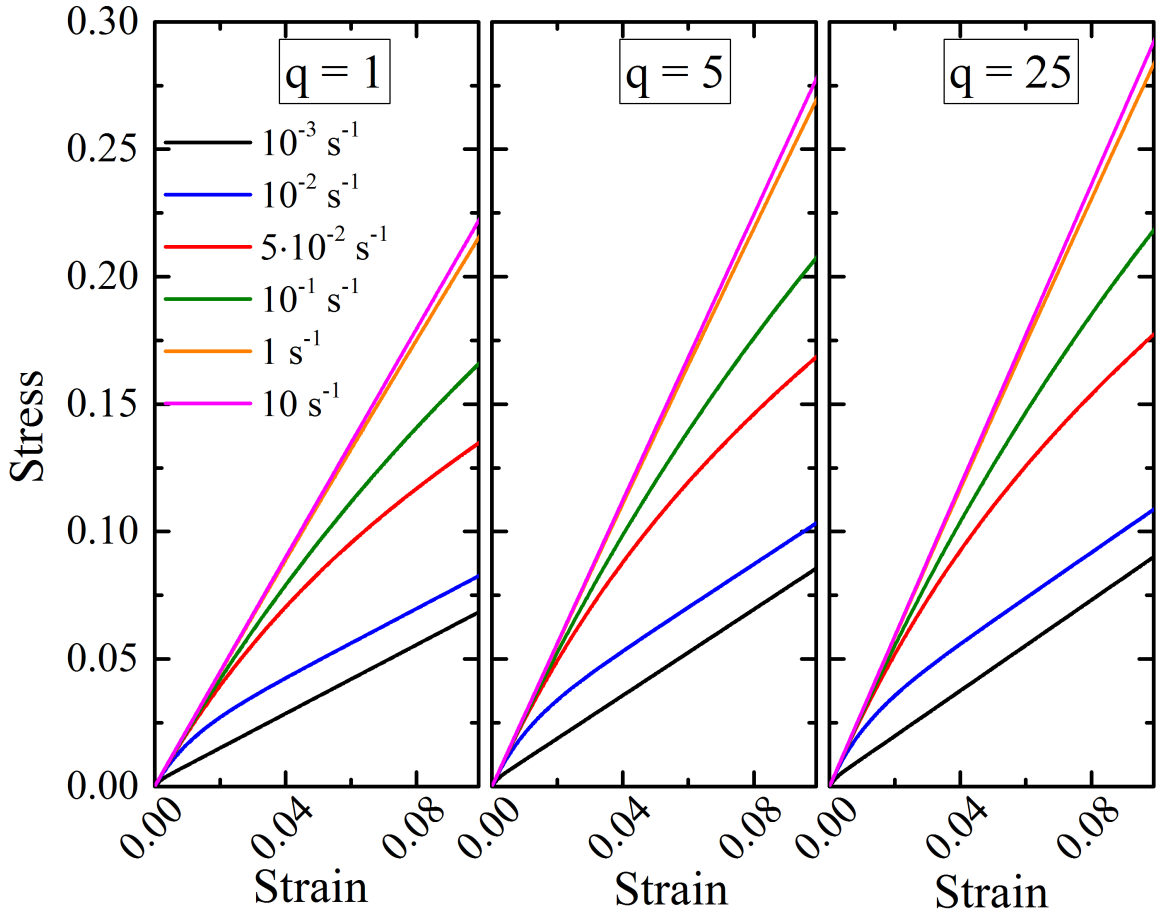

**Figure 6.** The modeled stress-strain curves, calculated according (24), for different values of instantaneous compressibility $q = \kappa/\mu$.

Figure 6 shows the corresponding stress-strain curves, calculated in accordance with Eq. (24), for different strain rates and different values of instantaneous compressibility $q$. One may see, that while compressibility affects the general behavior, the difference between the curves demonstrated in Figure 6a and Figure 6c for instantaneous compressibilities of 1 and 25 respectively, is mostly quantitative than qualitative. Therefore, we may conclude that introduction of the compressibility in the linear viscoelasticity is not able to explain the significant discrepancy between experimental and numerical data at the beginning stages of the loading (Figure 5b).

At the same time, there is no rule, which imposes the equality of the relaxation functions for volumetric and deviatoric parts of strain tensor [32]. Therefore, let us consider a more general case of the material with non-equal relaxation functions $D(t)$ and $H(t)$. Here we assume that both functions can be defined as one-term Prony serie, namely

$$D(t) = 1 - \gamma^D \left(1 - e^{-\frac{t}{\tau^D}}\right) \tag{26}$$

and

$$H(t) = 1 - \gamma^H \left(1 - e^{-\frac{t}{\tau^H}}\right). \tag{27}$$

It is easy to see that under this assumption for instantaneous and infinitely slow loading equations, the stress component $\sigma_{11}$ can be expressed as

$$\sigma_{11}(t) = \frac{9\mu\kappa}{3\kappa + \mu}\varepsilon_{11}(t) = \mu\frac{9q}{3q+1}\varepsilon_{11}(t) \tag{28}$$

and

$$\sigma_{11}(t) = \frac{9\mu(1-\gamma_D)\kappa(1-\gamma_H)}{3\kappa(1-\gamma_H) + \mu(1-\gamma_D)}\varepsilon_{11}(t) = \mu(1-\gamma_D)\frac{9q'}{3q'+1}\varepsilon_{11}(t), \tag{29}$$

where $q' = q(1-\gamma_H)/(1-\gamma_D)$ can be considered as the compressibility parameter for the infinitely slow loading.

Figure 7 shows the stress-strain curves obtained for $q = \frac{\kappa}{\mu} = 5$ and for different combinations of $q_c = q'/q$ and $\tau_c = \tau_H/\tau_D$, and compares them with the case of coincide relaxation functions, for which $q_c = \tau_c = 1$. The value of $q = 5$ corresponds to the material with Poisson ratio equals to 0.41 for instantaneous loading.

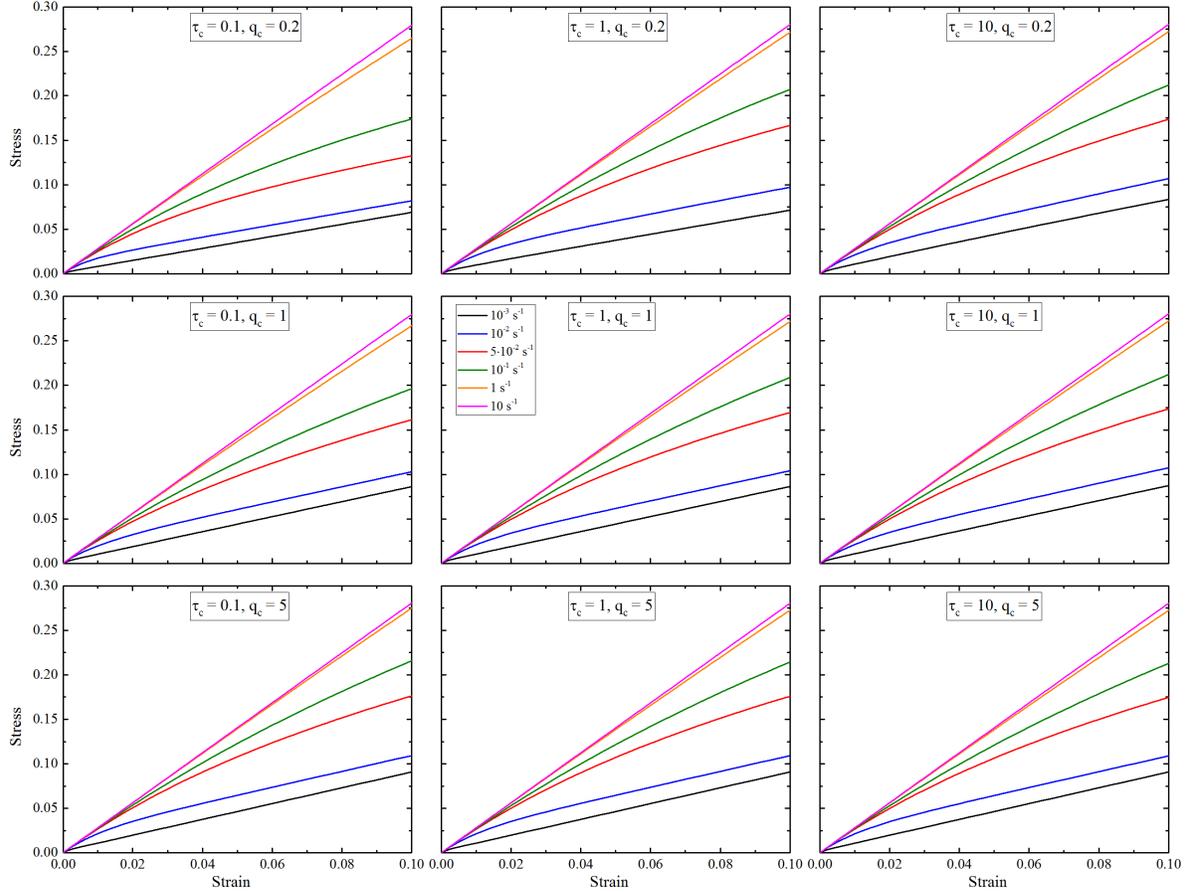

**Figure 7.** The modeled stress-strain curves under the assumption of (25) and (26) for different values of contrast in long-term shear modules $q_c$ and relaxation times $\tau_c$ between volumetric and deviatoric parts of the strain tensor.

As one may see from Figure 7, the difference between the relaxation times and coefficients of Prony series for volumetric and deviatoric parts of strain tensor affects the general mechanical behavior. However, the difference in the stress response is not significant. That allows us to assume that the reason of the inability of the QLV model to describe the rate-dependent behavior of stiff digital materials DM85 and DM95 is not caused by the incompressibility assumption. Probably, when the amount of stiff VW polymer exceeds some threshold value, deeper intrinsic processes, associated with polymeric chain interactions, result in more complicated mechanical response of these materials. Therefore, more complicated models are needed to describe the behavior of these DMs.